\documentclass[9pt,twocolumn,twoside]{opticastripped}
\setboolean{shortarticle}{false}
\setboolean{minireview}{false}

\title{Superpositions of up to six plane waves without electric-field interference}

\author[1,*]{K. C. van Kruining}
\author[2]{R. P. Cameron}
\author[3,4]{J. B. G\"otte}

\affil[1]{Max Planck Institute for the Physics of Complex Systems, N\"othnitzer Stra\ss{}e 38, 01187 Dresden, Germany}
\affil[2]{University of Strathclyde, 107 Rottenrow East,  G4 0NG Glasgow, United Kingdom}
\affil[3]{Nanjing University, 22 Hankou Road, Nanjing 210093, China}
\affil[4]{University of Glasgow, University Avenue, G12 8QQ, Glasgow, United Kingdom}

\affil[*]{Corresponding author: koen@pks.mpg.de}

\begin{abstract}
Superpositions of coherent light waves typically interfere. We present superpositions of up to six plane waves which defy this expectation by having a perfectly homogeneous mean square of the electric field. For many applications in optics these superpositions can be seen as having a homogeneous intensity. Our superpositions show interesting one- two- and three dimensional patterns in their helicity densities, including several that support bright regions of superchirality. Our superpositions might be used to write chiral patterns in certain materials and, conversely, such materials might be used as the basis of an ‘optical helicity camera’ capable of recording spatial variations in helicity.
\end{abstract}

\begin{document}

\maketitle

\section{Introduction}\label{sec:intro}
The electric and magnetic fields of a plane electromagnetic wave are orthogonal to each other and the direction of propagation. This suggests that the maximum number of waves with the same frequency that can be superposed without any interference is three. This can be done by choosing three waves travelling in mutually orthogonal directions and choosing all three polarisations orthogonal to each other. 

If one is content with only the mean square of the electric field being homogeneous without requiring that the mean square of the magnetic field also be homogeneous, larger superpositions are allowed. For many practical purposes, such superpositions can still be considered noninterfering, as it is the electric field that interacts most with matter, including fluorescent dyes, CCDs and the light-sensitive pigments in the human eye. The inhomogeneity in the magnetic field is relatively difficult to detect.
The helicity density, a quantity that indicates the handedness of the light \cite{TruebaRanada96, AfanasievStepanovski96, CameronBarnettYao12, CameronBarnett12, CameronBarnettYao14a}, is in general inhomogeneous for our noninterfering superpositions. It will vary in space in a pattern that is quite often, although not necessarily, periodic and resembles the intensity variations in optical lattices. There is enough freedom left in our superpositions to allow for a large variety of helicity lattices.

Some noninterfering superpositions show superchirality, an effect introduced by Tang \& Cohen \cite{TangCohen10, Hendryetal10, TangCohen11}. Superchiral light has regions where the helicity density is much higher than one would expect from the local mean square of the electric field. The key difference is that other superchiral superpositions exploit interference to create a region where the mean square of the electric field is weak, but the mean square of the magnetic field is not, allowing for the helicity density to become very large compared the squared local electric field \cite{TangCohen10, Rosales-GuzmanVolke-SepulvedaTorres12}. The helicity is actually quite small compared to the squared electric field outside of this `dark region'. Alternatively plasmonic resonators have been propopsed to generate high helicites close to a surface \cite{Schaferlingetal12, Schaferlingetal14, TianFangSun15, Schaferlingetal16, Krameretal17}. Noninterfering superpositions achieve superchirality in free space and in `bright regions' where the mean square of the electric field is not suppressed.

This article is structured as follows. In section~\ref{sec:construction} we show how to construct noninterfering superpositions. We then give several examples along with their helicity density patterns in section~\ref{sec:examples}. In section~\ref{sec:challenges} we estimate the residual inhomogeneity of the mean square of the electric field under small deviations from the exact required parameters. In section~\ref{sec:applications} we disuss the possibility of recording the helicity density patterns of our noninterfering superpositions using chirally sensitive liquid crystals and in section~\ref{sec:math} we discuss several open mathematical questions related to noninterfering superpositions.

\section{Construction of noninterfering superpositions and their helicity properties}\label{sec:construction}
In this paper we work in the classical domain  in free space and consider non-trivial superpositions of $N$ plane electromagnetic waves, each of which has the same angular frequency $\omega=ck$. For a superposition of $N$ waves the resulting electric and magnetic fields are
\begin{align}
\mathbf E=\mathrm{Re}\;\tilde{\mathbf E}=&\mathrm{Re}\left(\textstyle\sum_{j=1}^N\tilde{\mathbf E}_je^{i(\mathbf k_j\cdot\mathbf x-\omega t)}\right),\\
\mathbf H=\mathrm{Re}\;\tilde{\mathbf H}=&\mathrm{Re}\left(\textstyle\frac 1{\mu_0\omega}\sum_{j=1}^N\mathbf k_j\times\tilde{\mathbf E}_je^{i(\mathbf k_j\cdot\mathbf x-\omega t)}\right).\nonumber
\end{align}
By a non-trivial superposition we mean $\mathbf k_i\neq\mathbf k_j\;\forall\; i\neq j$ and $\tilde{\mathbf E}_j\neq 0\;\forall\; j $. The complex amplitudes $\tilde{\mathbf E}_j$ define the polarisations, amplitudes and phases of the waves. Because light's polarisation is transverse $\tilde{\mathbf E}_j\cdot\mathbf k_j=0\;\forall\;j$ applies.
\subsection*{Interference cancellation}
The mean square of the electric field is given by
\begin{multline}
\frac \omega{2\pi}\int_0^\frac{2\pi}\omega\mathbf{E\cdot E}d t=\frac 12\tilde{\mathbf E}\cdot\tilde{\mathbf E}^*\\
=\frac 12\left(\sum_{j=1}^N\tilde{\mathbf E}_j\cdot\tilde{\mathbf E}_j^*+\sum_{j=1}^N\sum_{l\neq j}\tilde{\mathbf E}_j\cdot\tilde{\mathbf E}_l^*e^{i(\mathbf k_j-\mathbf k_l)\cdot\mathbf x}\right).
\end{multline}
If this is to be homogeneous, the second sum must vanish. The most obvious way to achieve this is to choose the constituent plane waves such that no two interfere, in which case $\tilde{\mathbf E}_j\cdot\tilde{\mathbf E}_l^*=0\;\forall\; j\neq l$. This can be done for at most three plane waves because there are three orthogonal polarisation directions possible. In this paper we recognise that it is also possible, however, to allow multiple pairs of waves to interfere, provided the associated interference patterns cancel. To appreciate this, suppose that there exists within the superposition a pair $(i\neq j)$ of interfering waves, with the spatial periodicity of the associated interference pattern being dictated by the
wavevector difference $\mathbf k_j-\mathbf k_l$. If another pair $(j'\neq l')$ of interfering waves with the same wavector difference $\mathbf k_j-\mathbf k_l=\mathbf k_{j'}-\mathbf k_{l'}$ can be identified, giving an associated interference pattern with the same spatial periodicity, then the two interference patterns will cancel provided that $\tilde{\mathbf E}_j\cdot\tilde{\mathbf E}_l^*+\tilde{\mathbf E}_{j'}\cdot\tilde{\mathbf E}_{l'}^*=0$. The same reasoning applies for more than two pairs of interfering waves with the same wavevector difference. It is this trick that allows us to superpose more than three plane waves whilst keping the mean square of the electric field homogeneous.
\subsection*{Optical helicity and helicity lattices}
The definition of helicity density stems from plasma physics \cite{Woltjer58} and is known for arbitrary light fields in vacuum \cite{Calkin65, TruebaRanada96, AfanasievStepanovski96, CameronBarnettYao12, CameronBarnett12} and for an arbitrary medium with linear response \cite{vanKruiningGotte16}. As we are considering only monochromatic light fields we can use the simpler expression $\mathcal H=\frac 12\mathrm{Im}(\tilde{\mathbf E}\cdot \tilde{\mathbf H}^*)/c\omega$. The helicity density is bounded between $\pm\frac 1{4\omega}( \epsilon_0\tilde{\mathbf E}\cdot\tilde{\mathbf E}^*+\mu_0\tilde{\mathbf H}\cdot\tilde{\mathbf H}^*)$. For a superposition of $N$ plane waves one has
\begin{equation}
\mathcal H=\frac{- i}{4c\omega}\sum_{i,j=1}^N(\tilde{\mathbf E}_i\cdot \tilde{\mathbf H}_j^*-\tilde{\mathbf E}_j^*\cdot\tilde{ \mathbf H}_i)e^{i(\mathbf k_i-\mathbf k_j)\cdot\mathbf x}.
\end{equation}
When all waves are linearly polarised, the terms with $i=j$ are zero and only the `interference terms' remain. The vectors $\mathbf k_i-\mathbf k_j$ of all nonzero terms determine if the helicity density forms a lattice. If they are all linear combinations with integer coefficients of $\mathrm{dim}\!\left(\mathrm{span}\{\mathbf k_i-\mathbf k_j|\tilde{\mathbf E}_i\cdot \tilde{\mathbf H}_j^*-\tilde{\mathbf E}_j^*\cdot\tilde{ \mathbf H}_i\neq0\}\right)$  vectors, they form a lattice. If not, they form a less regular structure. Here span\{\,\} Is the space spanned by a set of vectors, dim(\,) is the dimension.

The superchirality threshold is $\sqrt{\epsilon_0 \mu_0}|\mathrm{Im}(\tilde{\mathbf E}\cdot \tilde{\mathbf H}^*)|>\epsilon_0 \tilde{\mathbf E}^*\cdot\tilde{\mathbf E}$ \cite{TangCohen10}. Because  $\tilde{\mathbf E}^*\cdot\tilde{\mathbf E}$ is homogeneous by design for all examples we will give, the occurrence of superchirality implies a stronger inequality, which we take as a sufficient condition for `bright superchirality': $\max\!\left(\sqrt{\epsilon_0 \mu_0}|\mathrm{Im}(\tilde{\mathbf E}\cdot \tilde{\mathbf H}^*)|\right)>\max\!\left(\epsilon_0 \tilde{\mathbf E}^*\cdot\tilde{\mathbf E}\right)$ (max=maximum). That is, the local helicity is large compared to the squared electric field anywhere. To the best of our knowledge, no monochromatic electromagnetic field with this property has ever been identified.
\section{Explicit examples}\label{sec:examples}
In this section our graphics of the helicity structure will always be $4\times4$ or $4\times 4\times 4$ wavelengths large unless stated otherwise and we use blue for negative helicity and red for positive helicity. 

We will show diagrams to illustrate the superpositions for which we plot the helicity density. In these diagrams grey arrows indicate wavevectors and electric field polarisations are indicated with yellow arrows. Green arrows to indicate the magnetic polarisations are included for reference as well. Mutually cancelling pairs of interference terms are indicated by red lines and interference terms contributing to the inhomogeneous helicity density are shown as black dashed lines. 

In every example we will give of a noninterfering superposition, there is some freedom left to change the amplitudes and propagation directions of the different waves. We will use $a_j$ to indicate free complex amplitudes which can take any nonzero value and $\theta_j$ and $\phi_j$ to indicate free angles where $0\le\theta_j<\pi$ and $0\le\phi_j<2\pi$, unless additional constraints are mentioned. An overall rotation of all wave- and polarisation vectors keeps a noninterfering superposition trivially noninterfering and we will not include this freedom in our parametrisations. Sometimes it is useful for convenient viewing of the helicity plots to include an overall rotation. We will indicate when we do this.

We observe the strongest superchirality when all free amplitiudes have roughly the same magnitude, although one can typically vary them by a factor of 2 or more before superchirality completely disappears. This is only a rule of thumb, because optimising a superposition for maximal superchirality is complicated and beyond the scope of this article.

\subsection*{Two waves}
Two plane waves can always be orthogonally polarised regardless of what their wavevectors are. One way to achieve this is to choose the polarisation of one wave to lie in the plane spanned by the wavevectors of both waves and the other one orthogonal to this plane \cite{CameronBarnettYao14a, CameronBarnettYao14b}. It is convenient to choose both waves to lie in the xy-plane, symmetrically with respect to the y-axis, see table~\ref{table:1}.

\begin{figure}\centering
\includegraphics[width=0.7\columnwidth]{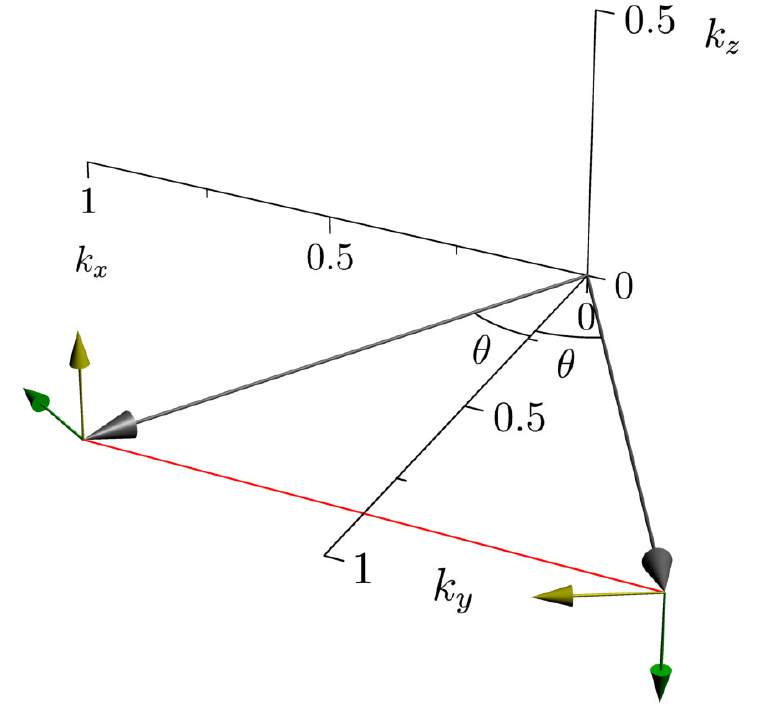}
\includegraphics[height=0.25\textheight]{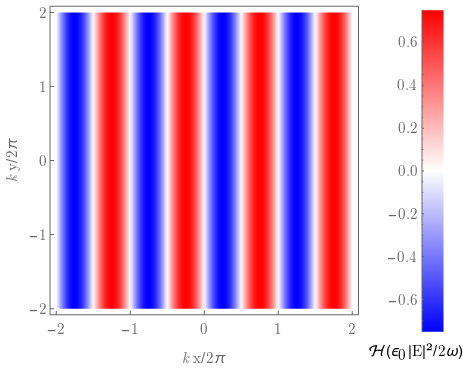}
\caption{The wave- and polarisation vectors for a noninterfering two-wave superposition (top) and its helicity pattern (bottom). For this superposition we chose both amplitudes equal and $\theta=\pi/6$.}\label{fig:2w}
\end{figure}

\begin{center}\begin{table}[h!]\begin{tabular}{|c|c|c|}\hline
$j$ & $\mathbf k_j$ & $\tilde{\mathbf E}_j$\\ \hline
1 & $[\sin\theta,\; \cos\theta,\;0]$ & $a_1[0,\; 0,\; 1]$\\ \hline
2 & $[{-}\sin\theta,\; \cos\theta,\; 0]$ & $a_2[\cos\theta,\;\sin\theta,\;0]$\\ \hline
\end{tabular}
\caption{A two-wave superposition with homogeneous mean square electric field.}\label{table:1}\end{table}\end{center}
For this configuration $\tilde{\mathbf E}\cdot\tilde{\mathbf E}^*$ and $\tilde{\mathbf H}\cdot\tilde{\mathbf H}^*$ are homogeneous. If one computes the helicity density, however, one does find a fringe structure similar to intensity interference fringes one would get if one chose the polarisations parallel, see Fig.~\ref{fig:2w}. 
\begin{equation}
\mathcal H=-\frac{\epsilon_0}\omega|a_1a_2^*| \cos^2\theta\sin\left(k_0\sin(2\theta)x+\arg(a_1a_2^*)\right).
\end{equation}
Because of the helicity-dependent force it exerts on chiral molecules \cite{Canaguier-Durandetal13, CameronBarnettYao14b} and birefringent microparticles \cite{MohantyRaoGupta05, Cipparroneetal10} this superposition has been studied in some detail. It has been found to act like a matter wave grating on chiral molecules \cite{CameronBarnettYao14c} similar to how a standing light wave can act like a grating for molecules in general \cite{HGUHNGBArndt09, EGArndtMT13, CameronGotteBarnettCotter16}.
\subsection*{Three waves}
\begin{figure}\centering
\includegraphics[height=0.25\textheight]{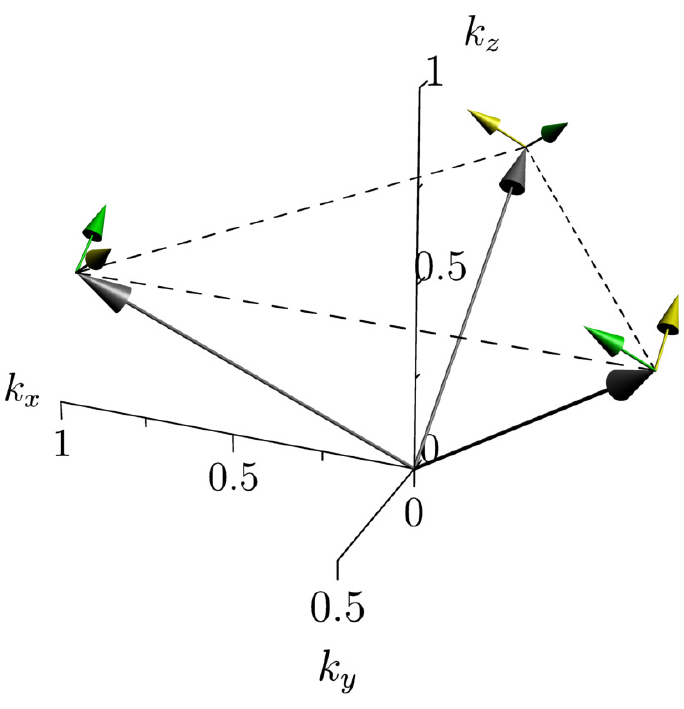}
\includegraphics[height=0.25\textheight]{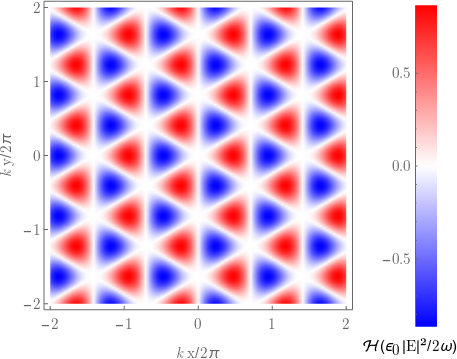}
\caption{The wave- and polarisation vectors for three noninterfering orthogonal waves ($\phi_1=\phi_2=\phi_3=0$), all amplitudes taken equal (top) and their helicity structure (bottom). For three plane waves the helicity interference terms always lie in a plane and thus form a two-dimensional helicity lattice. The wave vectors have been rotated to make the helicity lattice lie on the xy-plane.}\label{fig:3wb}
\end{figure}

Having three waves travelling in orthogonal directions and with orthogonal polarisations, yields both homogeneous $\tilde{\mathbf E}\cdot\tilde{\mathbf E}^*$ and $\tilde{\mathbf H}\cdot\tilde{\mathbf H}^*$. This setup already shows an interesting helicity structure, forming a triangular lattice, as is shown in Fig.~\ref{fig:3wb}. 
One is free to change the relative amplitudes. This will alter the shape of the positive and negative helicity regions within a unit cell, but keeps the lattice vectors fixed.

\begin{figure}\centering
\includegraphics[width=0.7\columnwidth]{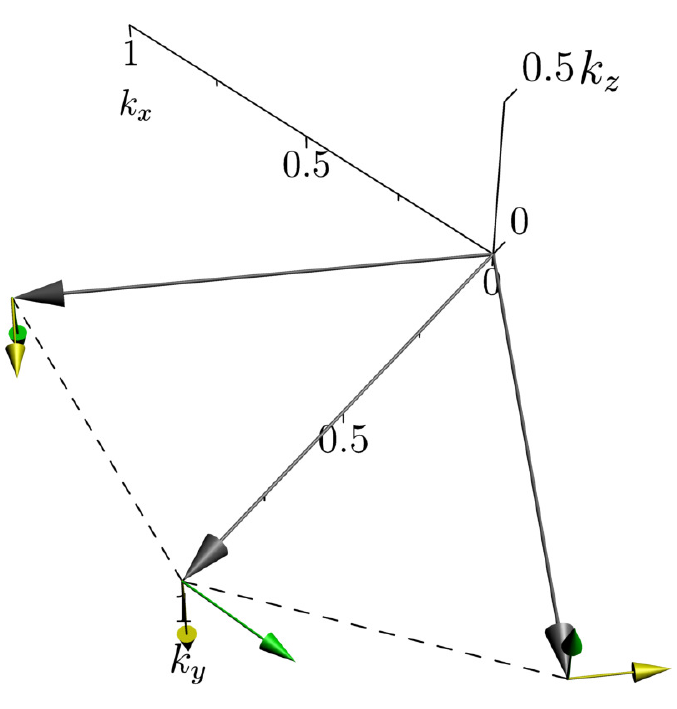}
\includegraphics[height=0.25\textheight]{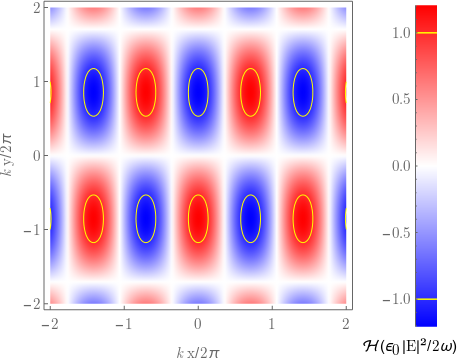}
\caption{The construction of a superchiral three-wave superposition (top). We have taken $\phi_1=0$, $\phi_2=\frac{7\pi}{4}$ and $\phi_3=\frac{3\pi}{2}$, rotated to make the helicity latrtice vectors parallel to the x- and y- axes. We took $a_2$ to be $\sqrt{2}$ times the amplitude of the other waves. The corresponding helicity structure (bottom) has superchiral regions (yellow ellipses) which extend for about half a wavelength in one direction and a quarter of a wavelength in the other.}\label{fig:3wc}
\end{figure}

If only $\tilde{\mathbf E}^*\cdot\tilde{\mathbf E}$ has to be constant, but $\tilde{\mathbf H}^*\cdot\tilde{\mathbf H}$ may vary, one can construct more superpositions by rotating the wavevectors of the three waves around axes given by their polarisation directions, see table~\ref{table:2}.
\begin{center}\begin{table}[h!]\begin{tabular}{|c|c|c|}\hline
$j$ & $\mathbf k_j$ & $\tilde{\mathbf E}_j$\\ \hline
1 & $[\cos\phi_1,\; 0,\; -\sin\phi_1]$ & $a_1[0,\; 1,\; 0]$\\ \hline
2 & $[{-}\sin\phi_ 2,\;  \cos\phi_2,\; 0]$ & $a_2[0,\;0,\; 1]$\\ \hline
3 &$[0,\; {-}\sin\phi_3,\;\cos\phi_3]$ & $a_3[1,\;0,\; 0]$\\ \hline
\end{tabular}\caption{A three-wave superposition with a homogeneous mean square electric field.}\label{table:2}\end{table}\end{center}
The additional freedom one gets from only requiring $\tilde{\mathbf E}^*\cdot\tilde{\mathbf E}$ to be homogeneous allows for a larger variety of helicity structures, including ones that are superchiral, as shown in Fig.~\ref{fig:3wc}. 
\subsection*{Four waves}
For superposing four waves without electric-field interference, one has to cancel a pair of interference terms against each other. One can take all wavevectors lying in a plane with their tips on the corners of a rectangle. If one takes one pair of waves polarised out of the plane and the other two polarised in plane there is only one nonzero pair of interference terms. One can then adjust the phases and amplitudes to make this pair of interference terms cancel. Taking all wavevectors in the xy-plane we parametrise the waves as shown in table~\ref{table:3}.

\begin{center}\begin{table}[h!]\begin{tabular}{|c|c|c|}\hline
$j$ & $\mathbf k_j$ & $\tilde{\mathbf E}_j$\\ \hline
1 & $[\cos \theta,\; \sin\theta,\; 0]$ & $a_1[0,\; 0,\; 1]$\\ \hline
2 & $[\cos \theta,\; {-}\sin\theta,\; 0]$ & $a_2[0,\; 0,\; 1]$\\ \hline
3 & $[{-}\cos\theta,\; \sin\theta,\; 0] $ &$\displaystyle-\frac{a_1a_2^*\mathrm{sgn}(\cos 2\theta)}{a_4^*\sqrt{|\cos 2\theta|}}[\sin\theta,\; \cos\theta,\; 0]$\\ \hline
4 & $[{-}\cos\theta,\; {-}\sin\theta,\; 0]$ & $\displaystyle\frac{a_4}{\sqrt{|\cos 2\theta|}}[{-}\sin\theta,\; \cos\theta,\; 0]$\\ \hline
\end{tabular}\caption{A four wave superposition with a homogeneous mean square electric field and all wave vecors lying in a plane. The angle $\theta$ can be anything between $0$ and $\frac \pi 2$ except $\frac \pi 4$.}\label{table:3}\end{table}\end{center}
The helicity density of this superposition is zero for $\theta<\frac \pi 4$ because both helicity terms cancel. For $\frac \pi 4<\theta<\frac \pi 2$ the helicity structure consists of sinusoidal fringes along the x-axis, a pattern that can already be achieved by superposing two waves \cite{MohantyRaoGupta05, CameronBarnettYao14c}. 

One can also take two waves lying in the xy-plane and polarised in the z-direction and two waves travelling in the yz-plane polarised in the x-direction. If both pairs of waves travel at the same relative angle, one can choose the amplitudes and phases such that their interference terms cancel, see table~\ref{table:4}. 
\begin{center}\begin{table}[h!]\begin{tabular}{|c|c|c|}\hline
$j$&$\mathbf k_j$ & $\tilde{\mathbf E}_j$\\ \hline
1 & $[\cos\theta,\; \sin\theta,\; 0]$ & $a_1[0,\; 0,\; 1]$\\ \hline
2 & $[\cos\theta,\; -\sin\theta,\; 0] $ & $a_2[0,\; 0,\; 1]$\\ \hline
3 & $[0,\;\sin\theta,\; \cos\theta] $& $\displaystyle-\frac{a_1a_2^*}{a_4^*}[1,\;0,\;0]$\\ \hline
4 & $[0,\;-\sin\theta,\; \cos\theta]$ & $a_4[1,\;0,\;0]$ \\ \hline
\end{tabular}\caption{A four wave superposition with a homogeneous mean square electric field and two waves in the xy-plane and two in the yz-plane. A fifth wave can be added to this superposition whilst keeping the mean square electric field homogeneous, see table~\ref{table:6}. }\label{table:4}\end{table}\end{center}
The helicity pattern of this superposition consists again of sinusoidal fringes, but this superposition allows for a fifth wave to be added whilst still keeping the mean square of the electric field constant. This five-wave superposition will be treated in the next subsection. 

\begin{figure}\centering
\includegraphics[width=0.7\columnwidth]{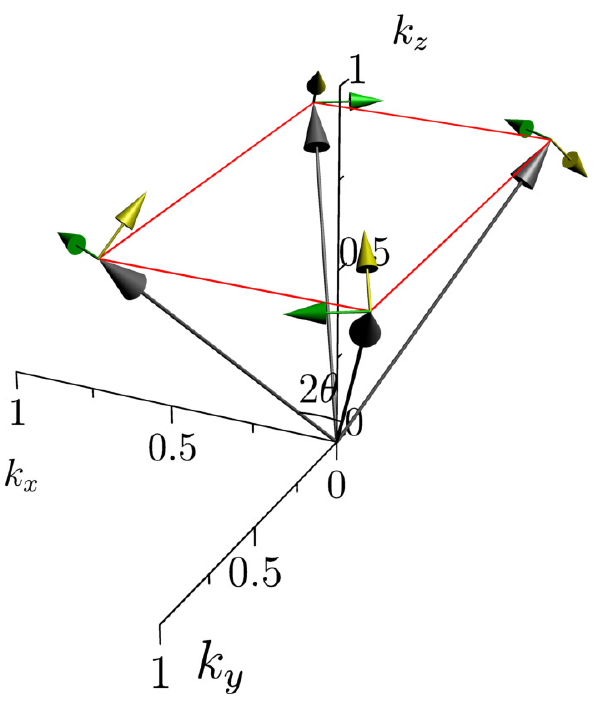}
\includegraphics[height=0.25\textheight]{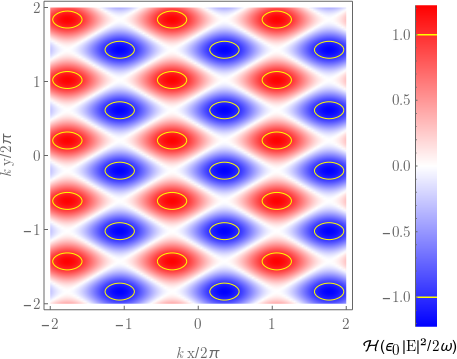}
\caption{A four-wave noninterfering superposition with two pairs of cancelling interference terms (top) and its helicity structure (bottom). For this superposition we chose $a_1=a_3$, $\theta=\frac\pi 6$ and $\Delta \phi=0$. Superchirality is typical for superpositions of this kind and occurs over broad parameter ranges.}\label{fig:4wb}
\end{figure}

There exists another four-wave superposition which involves the cancellation of two pairs of interference terms. It is constructed by taking a pair of waves travelling in orthogonal directions (we take them symmetric with respect to the z-axis) polarised in the plane spanned by the wavevectors. Then add a second copy of this pair rotated around the bisector of the first pair with an additional relative phase of $\pi$ between them, see table~\ref{table:5}
\begin{center}\begin{table}[h!]\begin{tabular}{|c|c|c|}\hline
$j$ & $\mathbf k_j$ & $\tilde{\mathbf E}_j$\\ \hline
1 & $\frac {\sqrt2}2[\cos \theta,\; \sin\theta,\; 1]$ & $a_1\frac {\sqrt2}2[{-}\cos \theta,\; {-}\sin\theta,\; 1]$\\ \hline
2 & $\frac {\sqrt2}2[{-}\cos \theta,\; {-}\sin\theta,\; 1]$ & $\displaystyle a_1^*\frac{a_3}{a_3^*}e^{i \Delta\phi}\textstyle\frac {\sqrt2}2[{-}\cos \theta,\; {-}\sin\theta,\; {-}1]$\\ \hline
3 & $\frac {\sqrt2}2[{-}\cos\theta,\; \sin\theta,\; 1] $ &$a_3\frac {\sqrt2}2[\sin\theta,\; {-}\cos\theta,\; 1]$\\ \hline
4 & $\frac {\sqrt2}2[\cos\theta,\; {-}\sin\theta,\; 1]$ & $a_3e^{i \Delta\phi}\frac {\sqrt2}2[{-}\sin\theta,\; \cos\theta,\; 1]$\\ \hline
\end{tabular}\caption{A superposition with homogeneous mean square electric field which can show interesting superchiral helicity lattices.}\label{table:5}\end{table}\end{center}
The helicity lattice formed by this superposition shows rhombs of positive and negative helicity arranged in a rectangular lattice, as is shown in Fig.~\ref{fig:4wb}. Superchirality occurs when $a_1$ and $a_3$ are of comparable magnitude.
\subsection*{Five waves}
The five-wave superposition we found is the only one we know about with a genuine three-dimensional helicity structure. It is constructed the following way. Take two plane waves propagating in the xy-plane polarised in the z-direction. Take two plane waves in the yz-plane propagating at the same relative angles and polarised in the x-direction. By choosing the amplitudes and phases of these waves right, one can cancel the interference terms between these waves, as explained in the previous subsection. Then one can add a fifth wave propagating in the xz-plane polarised in the y-direction. The parameters of this superposition are the shown in table~\ref{table:6}.
\begin{center}\begin{table}[h!]\begin{tabular}{|c|c|c|}\hline
$j$&$\mathbf k_j$ & $\tilde{\mathbf E}_j$\\ \hline
1 & $[\cos\theta,\; \sin\theta,\; 0]$ & $a_1[0,\; 0,\; 1]$\\ \hline
2 & $[\cos\theta,\; {-}\sin\theta,\; 0] $ & $a_2[0,\; 0,\; 1]$\\ \hline
3 & $[0,\;\sin\theta,\; \cos\theta] $& $\displaystyle-\frac{a_1a_2^*}{a_4^*}[1,\;0,\;0]$\\ \hline
4 & $[0,\;-\sin\theta,\; \cos\theta]$ & $a_4[1,\;0,\;0]$ \\ \hline
5 & $[\cos\phi,\;0,\;\sin\phi]$ & $a_5[0,\;1,\;0]$\\ \hline
\end{tabular}\caption{A five wave superposition with mean square electric field. This superposition yields three-dimensional helicity lattices.}\label{table:6}\end{table}\end{center}
\begin{figure}\centering
\includegraphics[height=0.7\columnwidth]{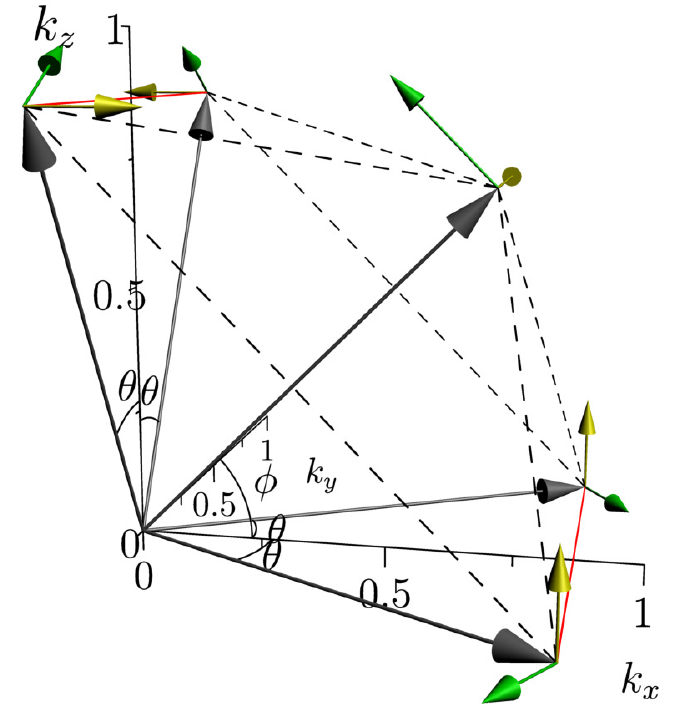}
\includegraphics[width=\columnwidth]{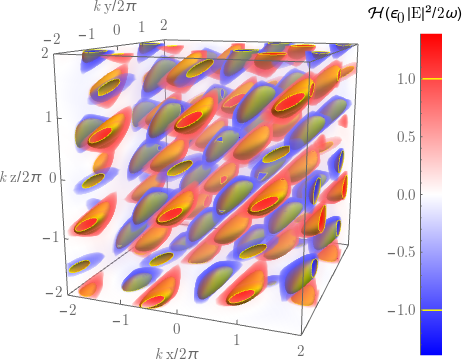}
\caption{An example of a five-wave noninterfering superposition (top) yielding a three-dimensional helicity lattice (bottom). For this superposition we took $\theta=\frac \pi 6$, $\phi=\frac \pi4$, $a_1=a_2=a_4=1$ and $a_5=\sqrt[4]8\sqrt{\cos\pi/6}$. The superchiral regions are enclosed by the yellow surfaces. At about half a wavelength in length they are surprisingly large.}\label{fig:5w}
\end{figure}
As one can see in Fig.~\ref{fig:5w}, the wavevectors lie on the corners of a (generically skewed) pyramid. The difference vectors between them form the sides of the pyramid. One can write all these difference vectors as linear combinations with integer coefficients of only three of them, choosing two that lie in the base and one connecting the base to the apex. Therefore the helicity structure of this superposition is generically periodic. For generic parameters the helicity lattice is monoclinic, with higher lattice symmetries for special parameters. If one chooses $\phi=\frac \pi4\vee \frac{5\pi}4$ the apex of the pyramid lies directly above the centre of the base, and the helicity lattice is orthorhombic. If the base of the pyramid is a square as well the helicity lattice is tetragonal. This is the case for $\theta=\arccos\frac 2{\sqrt 6}$.


For the special cases $\phi=\frac{3\pi}2+\mathrm{arctg}\left(1-\cos\theta\right)$ or $\phi=\pi-\mathrm{arctg}\left(1-\cos\theta\right)$ all difference vectors between the wavevectors lie in the same plane and the helicity structure is generically aperiodic, except if they all are rational linear combinations of only two of them, which happens for $\cos\theta/\sqrt 2\sin(\mathrm{arctg}\left(1-\cos\theta\right))\in\mathbb Q$. If this condition is met the helicity structure is a two-dimensional rectangular lattice, although the unit cell may be very large depending on the precise ratio. 

The large number of free parameters this superposition has allows one to construct superchiral helicity lattices with surprisingly pronounced ($\approx 1.4$ times the threshold value) and large (extending about half a wavelength in two directions) superchiral regions. We found that the most pronounced superchirality is achieved when the fifth wave points in roughly the same direction as the total momentum of the other four whereas having it point in the opposite direction attenuates the helicity modulations to far below the superchirality threshold. 

\subsection*{Six waves}
\begin{figure}\centering
\includegraphics[width=\columnwidth]{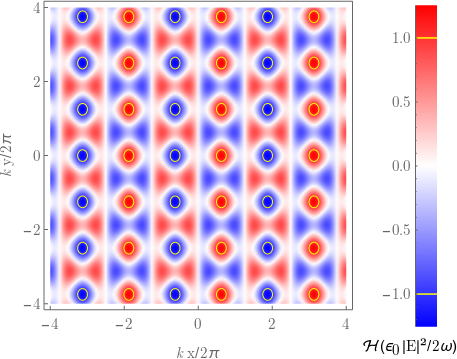}\vspace{7pt}
\includegraphics[width=\columnwidth]{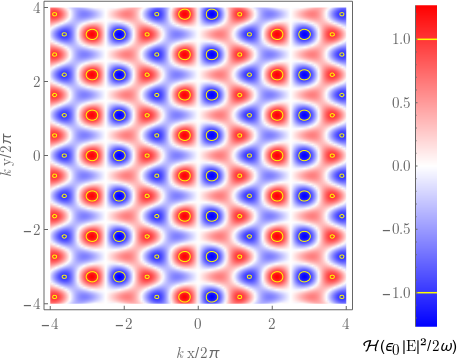}
\caption{A six wave helicity lattice with $\theta=\arccos-\frac 35$ and $a_1=a_2=a_3$ (top) and with $\theta=\arccos-\frac 25$ and $a_1=a_2=a_3$ (bottom). The size of these plots is $8\times8$ wavelengths.}\label{fig:6wb}
\end{figure}

One can superpose six waves by having three cancelling pairs of interference terms. The superposition is constructed by putting the six wavevectors on the corners of a hexagon. Three of them are polarised in the plane of the hexagon and three perpendicular to it. The plane waves in table~\ref{table:7} lead to a cancellation of all interference terms.
\begin{center}\begin{table}[h!]\begin{tabular}{|c|c|c|}\hline
$j$&$\mathbf k_j$ & $\tilde{\mathbf E}_j$\\ \hline
1 & $[1,\;0,\;0]$ & $a_1[0,\;0,\;1] $\\ \hline
2 & $[\cos \theta,\;\sin\theta,\; 0] $ &$ a_2[0,\;0,\;1]$\\ \hline
3 & $[\cos\theta,\; -\sin\theta,\; 0]$ & $a_3[0,\;0,\;1] $\\ \hline
4 & $[{-}1,\;0,\;0]$ & $\displaystyle-\frac{a_1^*\sqrt{|\cos 2\theta|}}{\cos\theta}[0,\;{-}1,\;0] $\\ \hline
5 & $ [{-}\cos\theta,\;{-}\sin\theta,\;0]$ & $\displaystyle\frac{a_2^*}{\sqrt{|\cos 2\theta|}}[\sin\theta,\;{-}\cos\theta,\;0]$\\ \hline
6 & $[{-}\cos\theta,\;\sin\theta,\;0]$ & $\displaystyle\frac{a_3^*}{\sqrt{|\cos 2\theta|}}[{-}\sin\theta,\;{-}\cos\theta,\;0]$\\ \hline
\end{tabular}\caption{A six wave superposition with homogeneous mean square electric field. Here $\theta$ limited to the intervals $\frac\pi4<\theta<\frac \pi2$ or $\frac\pi2<\theta<\frac{3\pi}4$. By varying the parameters, a large variety of two-dimensional helicity lattices are possible.}\label{table:7}\end{table}\end{center}
\begin{figure}\centering
\includegraphics[width=0.7\columnwidth]{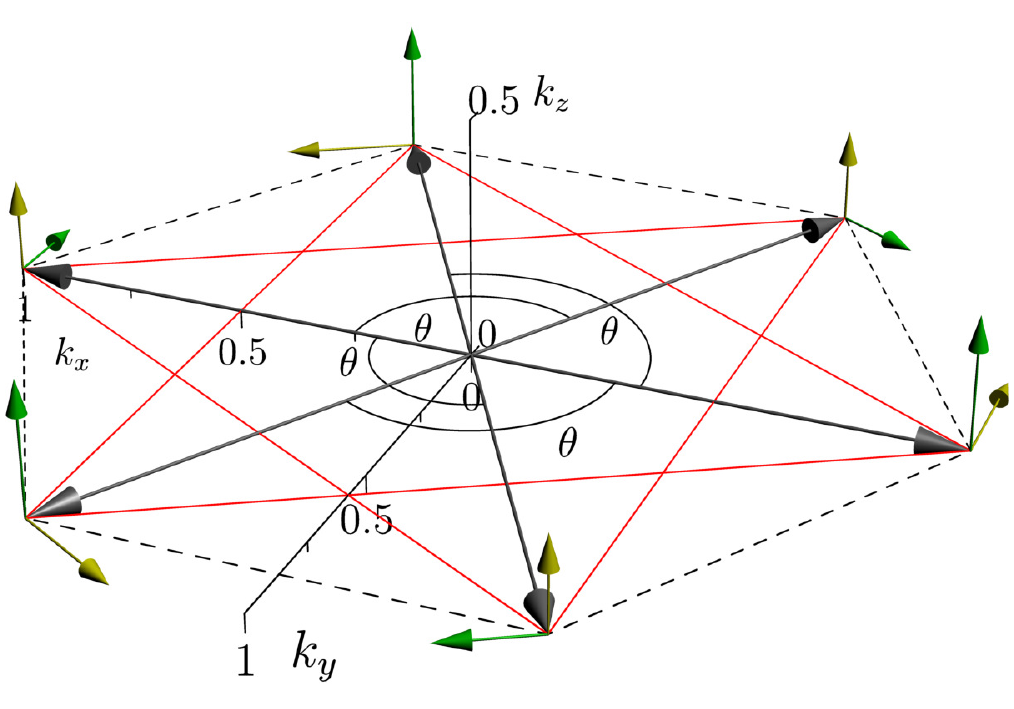}
\includegraphics[height=0.25\textheight]{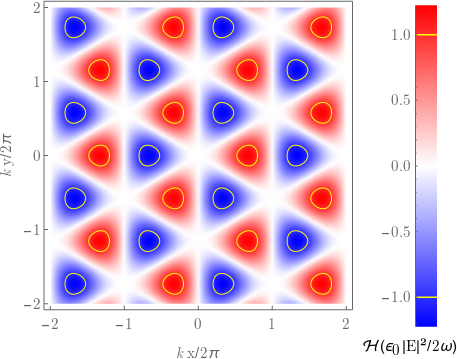}\vspace{7pt}
\includegraphics[height=0.25\textheight]{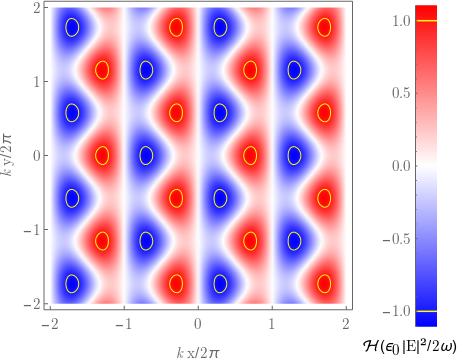}
\caption{The six wave superposition with $\theta=\frac{2\pi}3$ (top). This superposition requires three pairs of interference terms to cancel. For $\theta=\frac{2\pi}3$ the helicity forms a triangular lattice. Two examples are shown, one for $a_1=a_2=a_3$ (centre) and one for $a_1=\frac12 a_2=\frac 12a_3$ (bottom).}\label{fig:6wa}
\end{figure}
\begin{figure}
\includegraphics[width=0.5\textwidth]{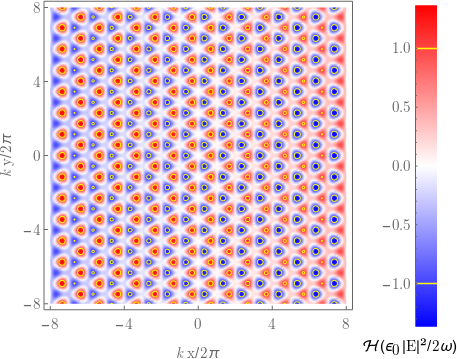}
\caption{The six-wave superposition for $\theta=\frac{2\pi}3-0.005$ and $a_1=a_2=a_3$. For this angle the helicity pattern is aperiodic in the x-dirextion. The size of this plot is $16\times16$ wavelegths.}\label{fig:6wmisaligned}
\end{figure}
Because all wavevectors lie in the same plane, the helicty structure is always two dimensional and with three sets of intereference terms contributing to the helicity structure (dashed lines in Fig.~\ref{fig:6wa}), it is not in general periodic in all directions, as one can see in Fig.~\ref{fig:6wmisaligned}. For the helicity structure to be periodic there have to exist two lattice vectors that have all $\mathbf k_i-\mathbf k_j$ contributing to the helicity structure as linear combinations with integer coefficients. This condition is equivalent to all $\mathbf k_i-\mathbf k_j$ being linear combinations with rational coefficients of two $\mathbf k_i-\mathbf k_j$ and is satisfied if $|\cos\theta|/(1-|\cos\theta|)\in\mathbb Q$. There exist an infinite number of angles that satisfy this condition, yielding an infinite number of different lattices, with unit cells being allowed to become arbitrarily large. In Fig.~\ref{fig:6wb} we give some examples of more complex helicity lattices. 

\section{The effects of small deviations from the exact parameters}\label{sec:challenges}
Although the superpositions from the previous section have some parametric freedom left, the homogeneity of $\tilde{\mathbf E}^*\cdot\tilde{\mathbf E}$ depends on the fine-tuning of at least some parameters. These parameters can be any of the parameters that specify a plane wave: amplitude, phase, polarisation or and propagation direction. The effects of misalignment in the propagation direction of the waves can be mitigated if the helicity lattice is of sufficiently small size and will be treated separately. Amplitude, phase and polarisation errors affect small lattices as well as big ones and can be treated in a unified picture.
\subsection*{Deviations in amplitude, phase and polarisation.}
Deviations in the amplitudes, phases and polarisations of the waves constituting a noninterfering superposition from their optimal values can be treated by writing the total electric field in the following way
\begin{equation}
\tilde{\mathbf E}=\sum_{j=1}^n (\tilde{\mathbf E}_j+\delta\tilde{\mathbf E}_j)e^{i \mathbf k_j\cdot\mathbf x}\quad\mbox{with}\quad \delta\tilde{\mathbf E_j}\cdot\mathbf k_j=0.
\end{equation}
The different components of $\delta \tilde{\mathbf E}_j$ represent the different parameters. The component parallel to and with the same complex phase as $\tilde{\mathbf E}_j$ causes a deviation in the amplitude, the component parallel and $\frac \pi2$ out of phase represents phase deviations, the component perpendicular to and with the same complex phase as $\tilde{\mathbf E}_j$ represents errors in the polarisation direction and the perpendicular component $\frac \pi2$ out of phase represents deviations in the ellipticity. Each of these components can cause residual interference in $\tilde{\mathbf E}^*\cdot\tilde{\mathbf E}$ at first order in $\delta\tilde{ \mathbf E}$. In general, the residual interference has the form:
\begin{equation}
\delta I_{\mathbf k\neq 0}=\frac 14\sum_{j\neq l}^n (\delta\tilde{ \mathbf E}_j^*\cdot\tilde{\mathbf E}_l+\tilde{\mathbf E}_j^*\cdot \delta \tilde{\mathbf E}_l)e^{i(\mathbf k_l-\mathbf k_j)\cdot\mathbf x}+O(\delta\tilde{\mathbf E}^2).
\end{equation}
As a measure of the quality of the superpositions we take the combined magnitude of all residual interference terms normalised by the homogeneous background field strength:
\begin{equation}
\frac{\sum_{j\neq l}^n\delta\tilde{ \mathbf E}_j^*\cdot \tilde{\mathbf E}_l+\tilde{\mathbf E}_j^*\cdot \delta \tilde{\mathbf E}_l}{\sum_{j=1}^n\tilde{\mathbf E}_j^*\cdot\tilde{\mathbf E}_j}=\frac{(n-1)\langle|\delta \tilde{\mathbf E}_j^*\cdot \tilde{\mathbf E}_l|\rangle_{jl}}{\langle \tilde{\mathbf E}_j^*\cdot \tilde{\mathbf E}_j\rangle_j}.
\end{equation}
Here $\langle\,\rangle_j$ denotes averaging over all waves and $\langle\,\rangle_{jl}$ denotes averaging over all pairs of different waves. We make the additional assumption that the expectation value of $|\delta\tilde{\mathbf E}_j|$ is independent of its orientation. That is, all parameters have equally big errors. If this is not the case, one can set the errors in all parameters equal to the least well controlled one as a worst-case estimate. Then, as a worst-case estimate we have $\langle|\delta \tilde{\mathbf E}_j^*\cdot \tilde{\mathbf E}_l|\rangle_{jl}\le \langle|\delta \tilde{\mathbf E}_j|\rangle_j\langle|\tilde{\mathbf E}_l|\rangle_l\frac 1{2\pi}\int_0^{2\pi}\cos\theta d \theta=\langle|\delta \tilde{\mathbf E}_j|\rangle_j\langle|\tilde{\mathbf E}_l|\rangle_l\frac 2\pi$ giving an estimate for the residual interference of
\begin{equation}
\frac{2(n-1)\langle|\delta \tilde{\mathbf E}_j|\rangle_j\langle|\tilde{\mathbf E}_l|\rangle_l}{\pi \langle \tilde{\mathbf E}_j^*\cdot \tilde{\mathbf E}_j\rangle_j}.
\end{equation}
From this equation one can see that the residual inhomogeneity of the helicity density is linear in polarisation, phase and polarisation errors and that superpositions of many waves are expected to have relatively larger helicity density inhomogeneities.
\subsection*{Deviations in the propagation direction}
Deviations in the propagation direction can be described by replacing $\mathbf k_j$ with $\mathbf k_j+\delta\mathbf k_j$ where $\mathbf k_j\cdot\delta\mathbf k_j=0$ to keep the frequency fixed. The effects of such deviations are twofold. First, if $\delta\mathbf k_j$ has a component parallel to $\tilde{\mathbf E}$, the light's polarisation must rotate accordingly to preserve transversality, introducing a change in the electric field of $\delta \tilde{\mathbf E}_j=\tilde{\mathbf E}_j\cdot \delta\mathbf k_j/|\mathbf k_j|$. Second, a pair of supposedly cancelling interference terms do not cancel exactly anymore, leading to a beating pattern in the field strength. Around the nodes the mean square of the electric field is relatively homogeneous and one can have a finite size low-interference region of size $L$ if
\begin{equation}
\delta \mathbf k_j<\frac{A_{\mathrm{max}}}{LA_{\mathrm{int}}},
\end{equation}
with $A_{\mathrm{max}}$ the maximal tolerable amplitude of $\tilde{\mathbf E}^*\cdot\tilde{\mathbf E}$-fluctuations and $A_{\mathrm{int}}$ the amplitude of a single interference term.
\section{Recording helicity patterns with liquid crystals}\label{sec:applications}
Over the past twenty years a rich variety of liquid crystal polymers were discovered that become chiral under illumination with circularly polarised light \cite{Nikolovaetal97, Nikolovaetal00, Iftimeetal00, Choietal06, Tejedoretal07, Veraetal07, BarrioTejedorOriol11, Vegaetal12, Daejongchem14}. The literature on the topic is far too broad to be covered here in full, but it suffices to note that the compounds that show this behaviour are either polymers with long light-absorbing side chains that twist themselves into helices under illumination \cite{Nikolovaetal97, Nikolovaetal00, Iftimeetal00, Tejedoretal07,  BarrioTejedorOriol11} or propeller shaped molecules that can stack themselves in either a left- or a right-handed helix \cite{Veraetal07, Vegaetal12, Daejongchem14}. The permanence of the induced chirality varies a lot betwen compounds. Some have their chirality erased by illumination with the opposite polarisation \cite{Iftimeetal00, Tejedoretal07} or by heating \cite{ Veraetal07}, others can have their chirality fixated \cite{Daejongchem14}. Light intensities used to achieve this chirality are on the order of tens or hundreds of miliwatt per square centimetre and illumination takes up to an hour \cite{Iftimeetal00, Tejedoretal07, Daejongchem14}. Several applications for these compounds have been tested, such as an optical polarisation switch \cite{Martinez-Ponceetal08} and chiral second harmonic generation \cite{Vegaetal12}. So far, these compounds were only used in combination with homogeneously polarised light. Helicity lattices can imprint an inhomogeneous chirality into a polymer film, making it possible to either use the polymer as `chiral' film to record the helicity structure of the light or using the light to write helicity-sensitive optical components into a polymer film. For example, the helicity patterns from Fig.~\ref{fig:6wb} can serve as arrays of chiral waveguides that guide light of one helicity only. The kind of polymer one would want for imprinting chiral structures is one that can chirally assemble under exposure with a helicity lattice and then have its supramolecular structure fixated by a process that works equally well on both enantiomers, yielding an imprint that remains stable at high temperatures and light intensities. 

\section{Some remarks on the mathematics of noninterfering superpositions}\label{sec:math}
We have constructed superpositions with homogeneous mean squared electric field of up to six plane waves. We do not know if there is an upper bound to the number of plane waves that can be superposed in this way, but we suspect there is, because the number of interference terms increases faster than the number of free parameters available. 

We also noticed that whenever we superpose four or more plane waves either $\tilde{\mathbf E}^*\cdot\tilde{\mathbf E}$ or $\tilde{\mathbf H}^*\cdot \tilde{\mathbf H}$ is inhomogeneous. We believe there cannot exist a superposition of four or more plane waves with both of them homogeneous but we did not find a rigorous proof of this conjecture. We have similarly been unable to prove that when superposing three or more waves either $\tilde{\mathbf E}^*\cdot\tilde{\mathbf E}$, $\tilde{\mathbf H}^*\cdot \tilde{\mathbf H}$ or $\tilde{\mathbf E}^*\cdot\tilde{\mathbf H}$ is inhomogeneous, alghough we suspect this to be the case as well.

Every noninterfering superposition of four or more waves we know about has all light waves linearly polarised and introducing elliptically polarised waves in any of them will lead to interference. This made us believe that all waves being linearly polarised is a requirement for every noninterfering superposition of four or more waves.

As far as we know, none of the above problems have been formulated before.
\section{Outlook}
Apart from being an optics curiosity, noninterfering superpositions of more than three waves raise new mathematical challenges and  provide new ways to probe and manipulate chiral matter. There is a large variety of possible helicity patterns, of which we only have shown a sample. For all helicity lattices the lattice spacing scales with the wavelength of the light and there is no size limit other than the technical ability to superpose multiple coherent beams of light at the desired wavelength. One can imagine X-ray helicity lattices with unit cells the size of atoms or radio wave lattices with unit cells bigger than a house.

We are surprised at how common `bright region' superchirality is among noninterfering superpositions. We expect that it can occur as well if $\tilde{\mathbf E}^*\cdot \tilde{\mathbf E}$ is inhomogeneous, making a systematic study of this effect go well beyond the scope of this article.  

\section*{Funding Information}
RPC's contribution was supported by the EPSRC (grant no. EP/M004694/1) and the Leverhulme trust (grant no. RPG-2017-048). JBG's contribution was supported by the national key research and development programme of China (contract no. 2017YFA0303700).
\section*{Acknowledgments}

We thank Ulf Saalmann for noticing the issue of robustness under small deviations from the optimal parameters.


\bibliography{../../literatuurlijst,../../boeken}


\end{document}